\documentclass[
    ,final            
  ]
  {aipproc}

\layoutstyle{6x9}

\begin{document}

\title{Experimental search for evidence of the three-nucleon force and 
a new analysis method}

\author{P. Th\"orngren Engblom}{
  address={ISV, Uppsala University, Box 535, S751 21 Uppsala, Sweden}
}
\author{H.O. Meyer}{
  address={Indiana University Cyclotron Facility, Bloomington, 
Indiana 47405, USA}
}
\author{B. von Przewoski}{
  address={Indiana University Cyclotron Facility, Bloomington, 
Indiana 47405, USA}
}
\author{J. Kuro\'s-\.Zo{\l}nierczuk}{
  address={Nuclear Theory Center, Indiana University, Bloomington, 
Indiana 47405, USA}
}
\author{T.J. Whitaker}{
  address={Indiana University Cyclotron Facility, Bloomington, 
Indiana 47405, USA}
}
\author{J. Doskow}{
  address={Indiana University Cyclotron Facility, Bloomington, 
Indiana 47405, USA}
}
\author{B. Lorentz}{
  address={Institut f\"ur Kernphysik, Forschungszentrum J\"ulich, 
J\"ulich, Germany}
}
\author{P.V. Pancella}{
  address={Western Michigan University, Kalamazoo, Michigan 49008, USA}
}
\author{R.E. Pollock}{
  address={Indiana University Cyclotron Facility, Bloomington, 
Indiana 47405, USA}
}
\author{F. Rathmann}{
  address={Institut f\"ur Kernphysik, Forschungszentrum J\"ulich, 
J\"ulich, Germany}
}
\author{T. Rinckel}{
  address={Indiana University Cyclotron Facility, Bloomington, 
Indiana 47405, USA}
}
\author{T. Wise}{
  address={University of Wisconsin, Madison, Wisconsin 53706, USA}
}
\author{H.~Wita{\l}a}{
  address={M. Smoluchowski Institute of Physics, Jagiellonian University,
30-059 Krak\'ow, Poland}
}
\author{J.~Golak}{
  address={M. Smoluchowski Institute of Physics, Jagiellonian University,
30-059 Krak\'ow, Poland}
}
\author{H.~Kamada}{
  address={Kyushu Institute of Technology, 
1-1 Sensucho, Tobata, Kitakyushu 804-8550, Japan}
}
\author{A.~Nogga}{
  address={Department of Physics, University of Arizona, Tucson, Arizona,
          85721, USA}
}
\author{R.~Skibi\'nski}{
  address={M. Smoluchowski Institute of Physics, Jagiellonian University,
30-059 Krak\'ow, Poland}
}
\begin{abstract}
A research program with the aim of investigating the spin dependence
of the three-nucleon continuum in $\overrightarrow p \overrightarrow d$
collisions at intermediate energies was carried out at IUCF using the 
Polarized INternal Target EXperiments (PINTEX) facility. In the elastic 
scattering experiment at 135 and 200 MeV proton beam energies a total of 
15 independent spin observables were obtained.
The breakup experiment was done with a vector and tensor polarized 
deuteron beam of 270 MeV and an internal polarized hydrogen gas target.
We developed a novel technique for the analysis of the breakup 
observables, {\it the sampling method}. The new approach takes into account 
acceptance and non-uniformities of detection efficiencies and is suitable 
for any kinematically complete experiment with three particles in the final 
state.
\end{abstract}
\maketitle
\subsection{Introduction}
The goal of finding experimental evidence for a three-nucleon force,
and especially to map out its spin dependence has to include the major 
part of possible observables 
in order to be conclusive. A prerequisite for the interpretation of the 
data is the possibility to compare the results to computationally exact 
solutions of the three-nucleon system. 
To meet this theoretical challenge the Cracow-Bochum group 
has performed Faddeev calculations up to the pion production 
threshold \cite{HW-FADD:PRC63}. 
Using modern high-precision nucleon-nucleon (NN) potentials 
in combination with models for three-nucleon forces (3NF), 
parameter-free predictions for scattering observables can be obtained. 
Deviations from NN behaviour and agreement with calculations incorporating 
a 3NF model is thus considered to be the signature for the existence of 3NFs.
Considerable experimental activity over the last few years has been aimed 
primarily at measuring vector and tensor analyzing powers in pd elastic 
scattering, recently also several transfer coefficients were reported
(\cite{SEK:proc} and references therein). 

\subsection{The PINTEX pd elastic scattering experiment}
\begin{figure}[htb]
\centerline{
 \includegraphics[width=19cm,bb=-50 100 650 710,clip=true]{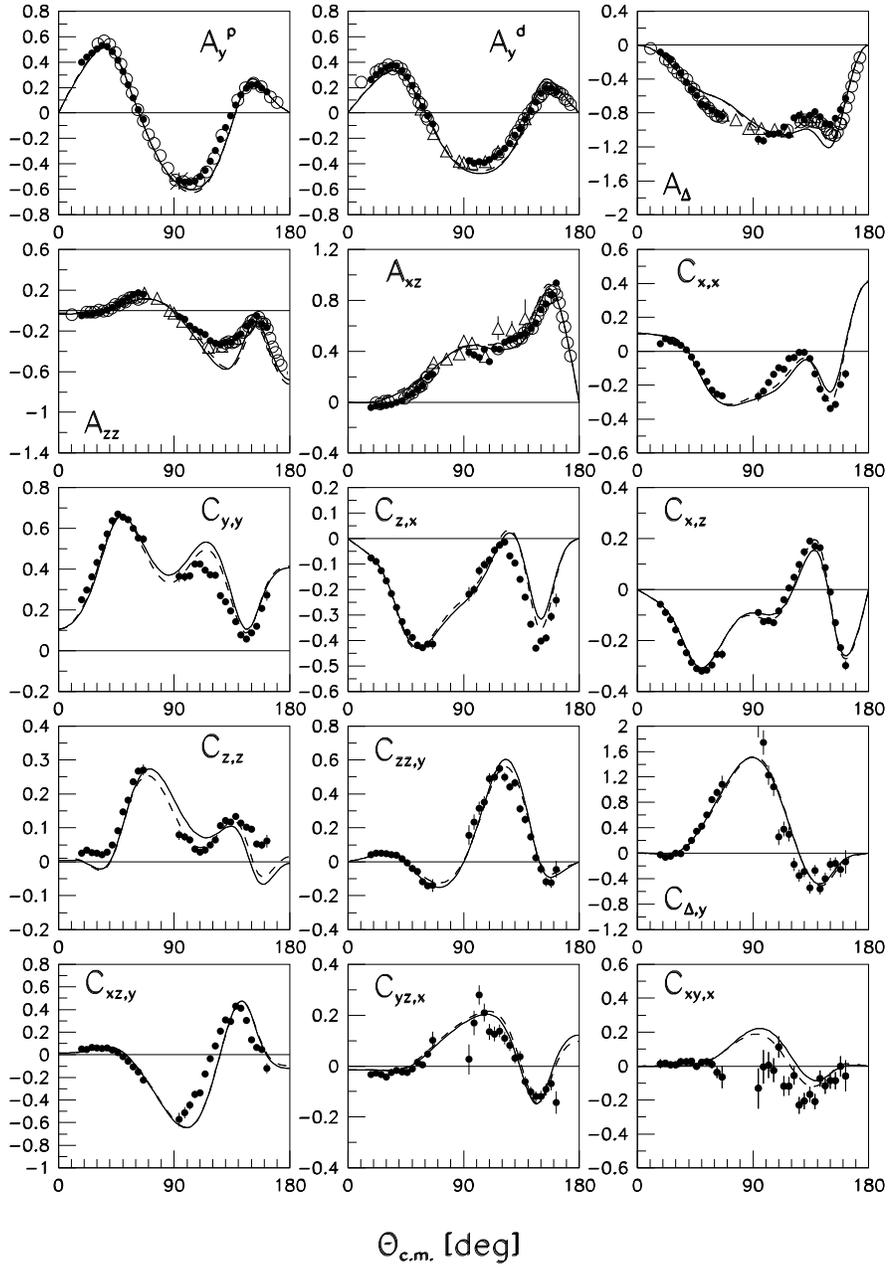}
 \caption{Spin observables at 135 MeV from the present experiment 
(full circles) together with previously existing data (empty markers) 
and theoretical  predictions from NN potentials: CDBonn (full line) 
\cite{RM-CDBonn:PRC63}, AV18 \cite{AV18:95}(dashed line). Figure taken 
from \cite{BvP:PRC04}.} 
\label{fig:135ce80}
}
\end{figure}

In a Cooler storage ring experiment a polarized proton beam at 135 and 
200 MeV was used in conjunction with a vector and 
tensor polarized deuterium gas target. The advanced options of the facility 
to control and change the spin alignment axis of both protons and deuterons 
allowed all analyzing powers and 10 of 12 possible spin correlation 
coefficients to be measured \cite{BvP:PRC04}.
The results at 135 MeV are shown in Fig.\ref{fig:135ce80}. The agreement of
the data with NN potential predictions is fairly good. The effects
of the inclusion of a 3NF was studied and did not improve the agreement
between theory and data in any consistent way with respect to scattering 
angle, energy or particular 3NF model 
(see figures 10-12 of ref \cite{BvP:PRC04})

\begin{figure}[htb]
\centerline{
 \includegraphics[width=5cm,bb=50 150 600 650,clip=true]{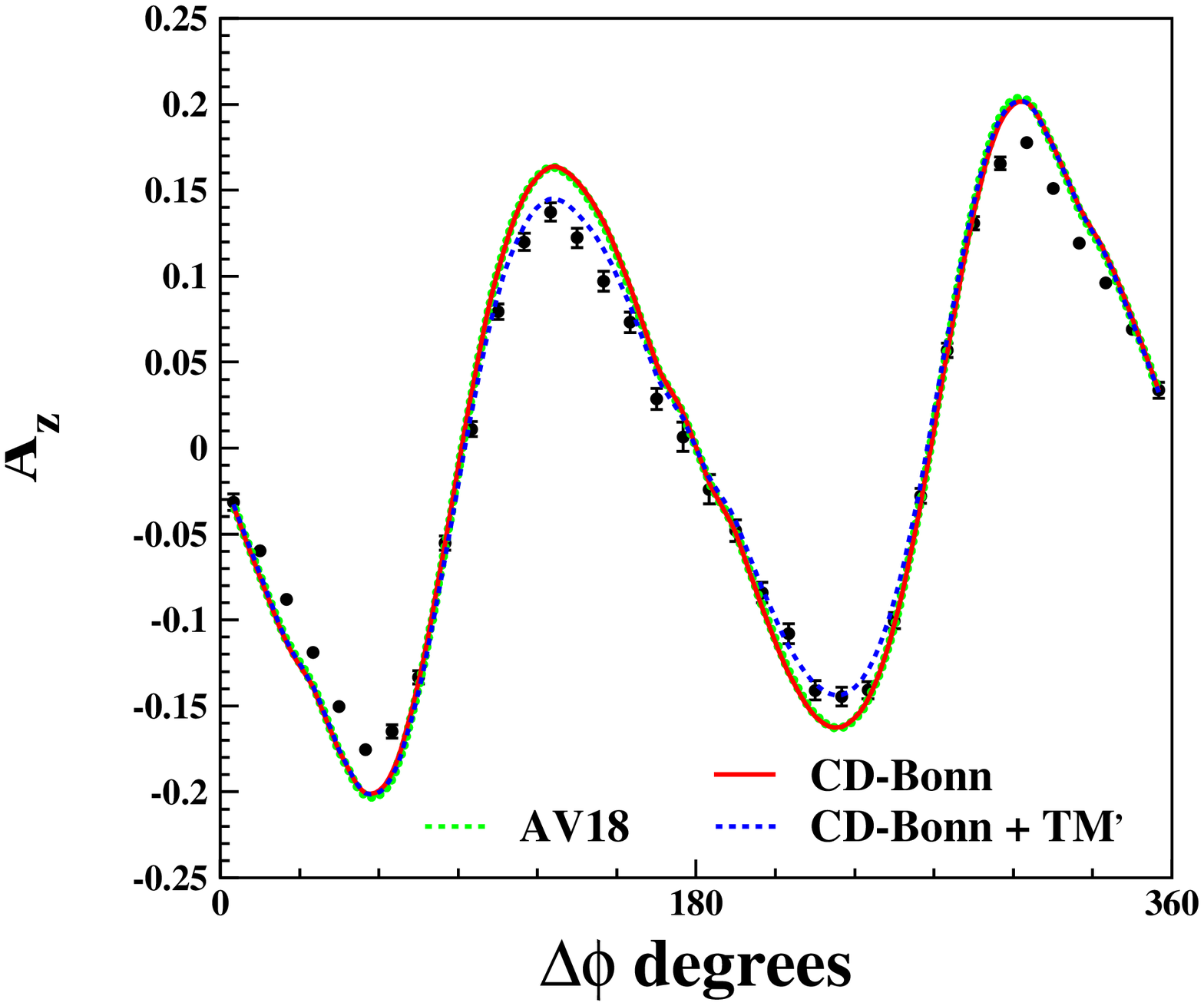}
\hspace{0cm}
\includegraphics[width=5cm,bb=50 150 600 650,clip=true]{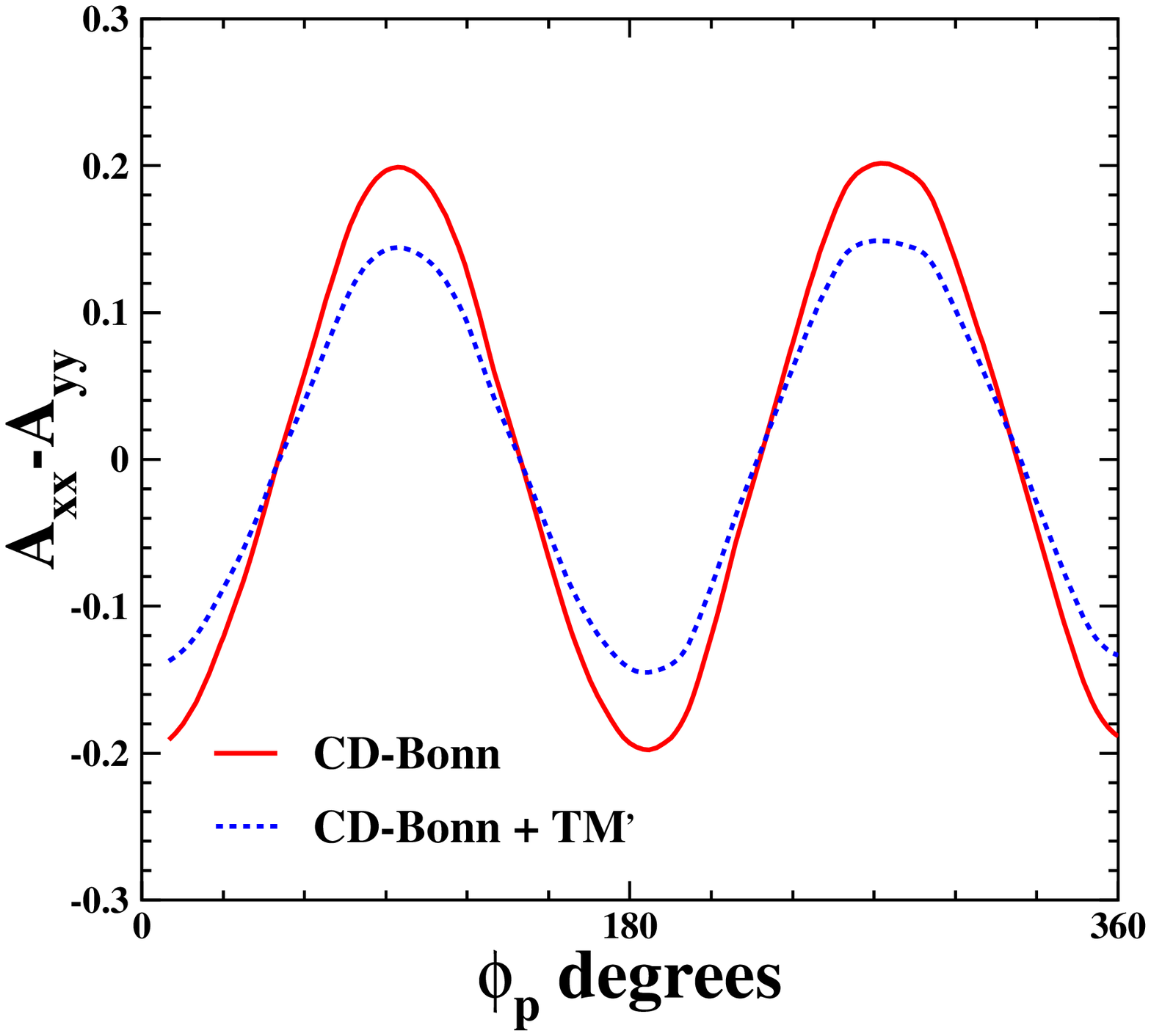}
\hspace{0cm}
 \includegraphics[width=5cm,bb=50 150 600 650,clip=true]{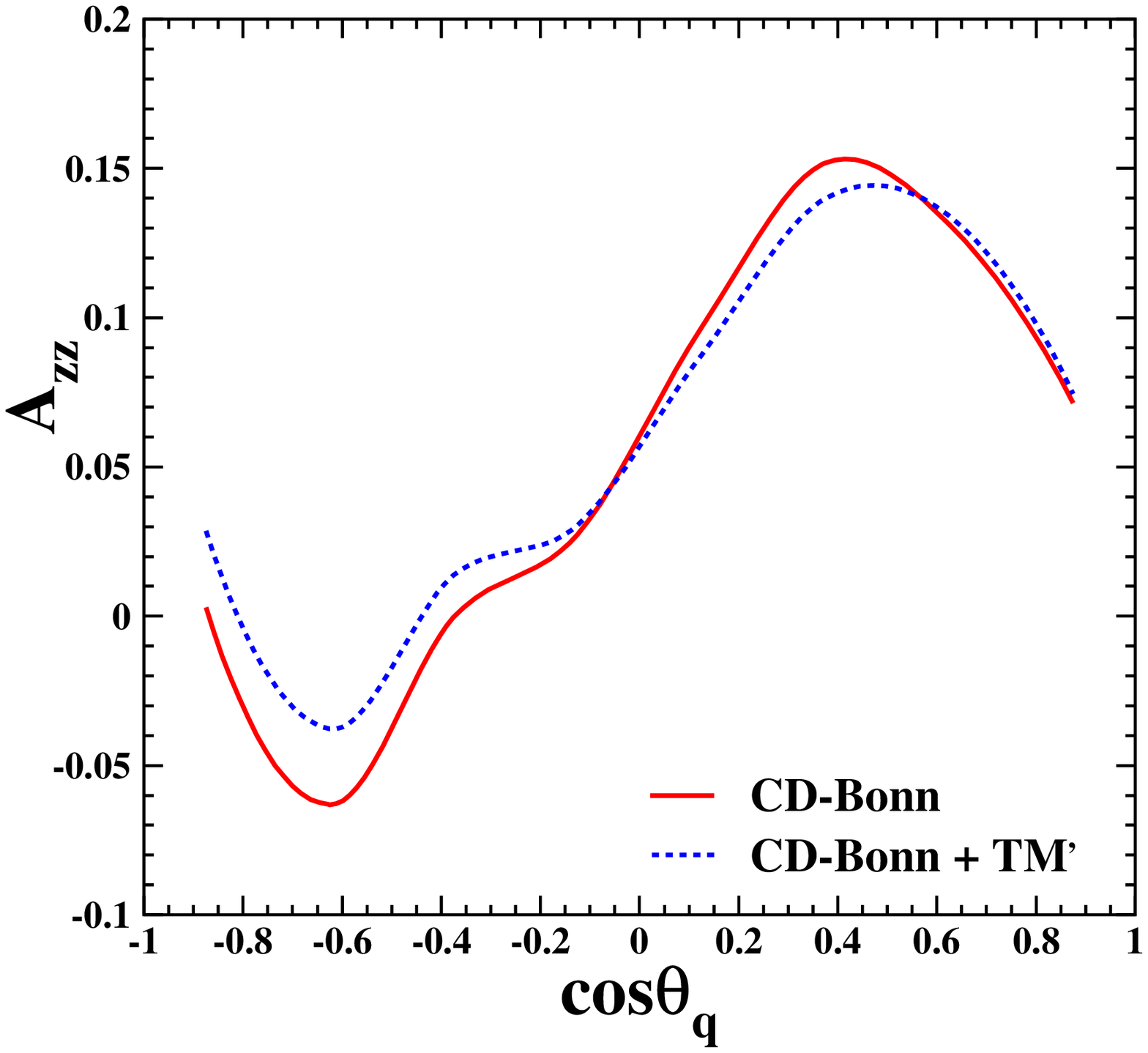}
}
\label{fig:tensorobs}
\caption{{\it Left:} Measured $A_z$ as a function of 
$\Delta \phi= \phi_p-\phi_q$, (where $p$ and $q$ are the jacobi momenta
in the cm), compared to NN and 3NF calculations.
{\it Middle \& Right:}
Predicted 3NF sensitivity for $A_{xx}-A_{yy}$ and $A_{zz}$ 
for the present experimental setup obtained by {\it the sampling method} 
and interpolation. As input phase space coordinates for the interpolation 
were used: a data sample (middle panel) and phase space distributed 
montecarlo generated events (right panel).}
\end{figure}
\subsection{The dp breakup channel}
The analysis of the dp breakup observables has been somewhat restricted 
by the fact that comparison to theory could only be done over 
limited phase space regions as a function of the kinematically allowed  
locus in the plane of the energies of the two detected nucleons (the 
so called S-curve). In order to overcome this deficiency and make use of the 
relatively large acceptance of the PINTEX facility we developed 
a new analysis technique; {\it the sampling method} \cite{KTM:FBS04}. 

For a three-particle final state to be kinematically fully determined, 
five parameters are required. When extracting observables one has to 
choose which independent variable to use 
and what regions of phase space to integrate over. The acceptance and 
any significant efficiency variation has to be well known, most often this
is accomplished by advanced monte carlo simulations. However, with the 
complete kinematical information of an event as input, the theoretical 
prediction for the sought observable can be calculated for that particular 
event. A given experimental data set ($\gamma$), containing the phase space 
points ($x_i$) used as input to a theoretical model calculation, can then be 
compared directly to theory by taking the mean of the calculated theoretical 
predicted values:
$O^{th}(\gamma )=\frac{\sum  O^{th}(x_i)}{ N(\gamma )}=<O^{th}>$,
where $N(\gamma)$ is the total number of events. The statistical error 
of the theoretical prediction that appear from the randomness of the sampling,
is given by the standard deviation. 
In case the theoretical computations are very time consuming, as when the
Faddeev amplitudes are needed, we adopt multidimensional linear interpolation
on a grid of precalculated stored theoretical values, for details see 
\cite{KTM:FBS04}. The {\it sampling method using a grid} is particularly 
useful in planning experiments and in governing a complicated data analysis.
See Fig. \ref{fig:tensorobs} for examples of theoretical predictions 
for the tensor analyzing powers in dp breakup as measured by our 
experimental setup.

For the experiment the deuteron beam was prepared in five different 
polarization states, unpolarized, two pure tensor($\pm$), and two 
mixed vector($\pm$) and tensor$+$ states. The target was a thin (25$\mu m$) 
Teflon-coated aluminum storage cell filled by an atomic beam of polarized 
hydrogen atoms. The four-momenta of the two outgoing protons were 
measured in a forward detector stack subtending a polar angle of 45 
degrees \cite{RT:NIM00}.

The {\it sampling method} was applied to the
analysis of axial observables in breakup \cite{HOM:PRL04}. 
The experiment was motivated by the prediction that axial observables 
may exhibit enhanced sensitivity to the 3NF due to a particular kind 
of spin operators present in a 3NF but absent in NN interactions 
\cite{LDK:PRL94}. Axial observables vanish for coplanar events 
since parity is conserved, whereas they may be non-zero in any other 
configuration of a breakup reaction. It was found that the agreement
with theory was not improved by including a 3NF in the calculations 
and thus the expectations concerning axial observables was not validated
by the experiment.
\subsection{Summary}
Large values of axial observables in $\overrightarrow d \overrightarrow p$ 
breakup were reported for the first time using a new analysis technique, 
applicable to any three-particle final state that is fully determined. 
The analysis of the breakup tensor analyzing powers is in progress with 
the objective to compare with theoretical predictions extracted from an 
NN potential at fourth order of ChPT \cite{RM-N3LO:PRC68}. The 
currently available 3NF models are not confirmed by our measurements 
of pd elastic and breakup reactions.
\begin{theacknowledgments}
The authors wish to thank the IUCF Operations Group for their enthusiastic
work at all hours. The work has been done under NFS Grant No. PHY-0100348, 
DOE Grant No. FG0288ER40438 and Grant No. Dnr. 629-2001-3868 from the Swedish
Research Council.
\end{theacknowledgments}

\end{document}